\documentclass[aps, prl, reprint,10pt,longbibliography]{revtex4-1}

\usepackage[caption=false]{subfig}

\usepackage{amsmath, amssymb, amsfonts, amsthm, bm, mathrsfs, bbm, enumerate}
\usepackage{graphicx}
\usepackage{epstopdf}
\usepackage{color}
\usepackage[usenames,dvipsnames]{xcolor}
\usepackage[citecolor=blue, colorlinks=true, urlcolor=blue, linkcolor=blue]{hyperref}

\usepackage{feynmp-auto} 

\newcommand{\eqw}[1]{(\ref{#1})}
\newcommand{\eq}[1]{Eq.~(\ref{#1})}

\newcommand{\fig}[1]{Fig.\thinspace{}\ref{#1}}

\newcommand{\fc}[1]{({#1})}
\newcommand{\figc}[2]{Fig.\thinspace{}\ref{#1}\thinspace{}\fc{#2}}

\begin{document}

\title{Periodically Driven Sachdev-Ye-Kitaev Models}

\author{Clemens Kuhlenkamp}
\affiliation{Department of Physics and Institute for Advanced Study, Technical University of Munich, 85748 Garching, Germany}
\affiliation{Munich Center for Quantum Science and Technology (MCQST), Schellingstr. 4, D-80799 M{\"u}nchen, Germany}

\author{Michael Knap}
\affiliation{Department of Physics and Institute for Advanced Study, Technical University of Munich, 85748 Garching, Germany}
\affiliation{Munich Center for Quantum Science and Technology (MCQST), Schellingstr. 4, D-80799 M{\"u}nchen, Germany}

\begin{abstract}
Periodically driven quantum matter can realize exotic dynamical phases. In order to understand how ubiquitous and robust these phases are, it is pertinent to investigate the heating dynamics of generic interacting quantum systems. Here we study the thermalization in a periodically driven generalized Sachdev-Ye-Kitaev (SYK) model, which realizes a crossover from a heavy Fermi liquid (FL) to a non-Fermi liquid (NFL) at a tunable energy scale. Developing an exact field theoretic approach, we determine two distinct regimes in the heating dynamics. While the NFL heats exponentially and thermalizes rapidly, we report that the presence of quasiparticles in the heavy FL obstructs heating and thermalization over comparatively long timescales. Prethermal high-frequency dynamics and possible experimental realizations of non-equilibrium SYK physics are discussed as well.
\end{abstract}

\date{\today}

\pacs{
}

\maketitle

Coherent periodic driving emerges as a fascinating new tool to induce novel properties both in synthetic quantum systems and in the solid state. Examples include the manipulation of topological band structures~\cite{oka, kitagawa, lindner, gedik2013, RechtsmanNature2013, aidelsburger2013, Miyake2013, Jotzu2014}, the realization of light-induced ordered states~\cite{fausti, mitrano, Subedi2014, Sentef2014, Knap2015, Babadi, Murakami}, and driving dynamical transitions from many-body localized to ergodic phases~\cite{bordia_periodically_2016, Ponte2015, Achilleas15, gopalakrishnan_regimes_2016}. While it may be possible to stabilize such exotic driven quantum states in an intermediate prethermal regime~\cite{BukovGopalakrishnan2015, CanoviKollar2015, LindnerBerg2016, Weidinger, Else_Prethermal_2017} or by adding disorder~\cite{bordia_periodically_2016}, the generic fate of an isolated periodically driven system is that it absorbs energy from the drive and heats toward an infinite temperature state, provided that the driving frequency is low enough~\cite{rosch}. By contrast, for high driving frequency this absorption is inefficient as it requires large rearrangements in the many-body state~\cite{AbaninHeating, HeatingBound, kuwahara2016, Else_Prethermal_2017}. To elucidate this interplay, it would be valuable to find a model that captures the effect of strong interactions yet can be solved in the thermodynamic limit.

\begin{figure}[t]
	\def\svgwidth{0.47\textwidth}
	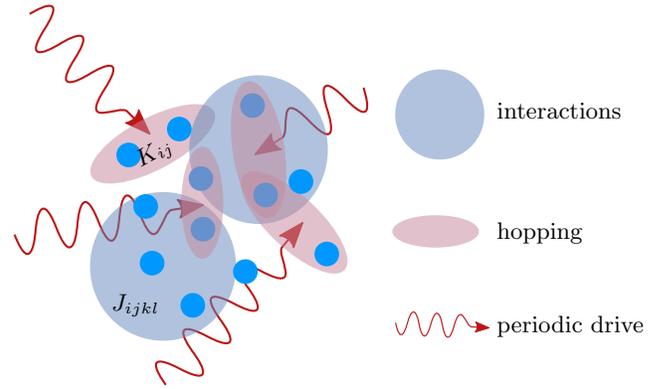
	\caption{\textbf{A periodically driven Sachdev-Ye-Kitaev model.} Illustration of a generalized SYK model with periodically modulated hopping terms (red ellipses) and a static interaction Hamiltonian (blue ellipses). The drive of the hopping is illustrated by the red wiggly lines.}
\label{fig:schematic}
\end{figure}

In this work, we propose that the generalized Sachde-Ye-Kitaev (SYK) model~\cite{SachdevYe, KitaevTalk} is a prototypical model for fast heating in periodically driven quantum systems. The SYK model was originally introduced as a solvable model for a non-Fermi liquid~\cite{SachdevYe, ParcolletGeorgesNFL}. Recently, work initiated by Kitaev~\cite{KitaevTalk, Polchinski, Maldacena2016Remarks, Bagrets_SYK_2016} studied quantum dynamics, chaos, and information scrambling in the SYK model and showed that it saturates bounds for the operator growth in out-of-time ordered correlation functions~\cite{Maldacena2016Remarks}. Remarkably, generalizations of SYK models feature a crossover from a heavy Fermi liquid (FL) to a non-Fermi liquid (NFL) and can be solved exactly~\cite{balents,Pengfei2017,altman_syk}. By periodically driving these models we study the heating dynamics of a heavy FL and a NFL in the thermodynamic limit; see \fig{fig:schematic} for an illustration. While we find the NFL to rapidly thermalize, the existence of well-defined quasiparticles dramatically slows down the full thermalization of the heavy FL. 
Furthermore, heating can be suppressed in both cases by driving with sufficiently high frequencies. The system then enters a prethermal regime characterized by an effective temperature which is stable over long times. We also explore a possible experimental realization of our system by measuring the absorption of a THz laser that impinges on an irregularly shaped and perforated graphene flake~\cite{ChenGrapheneFlake}.

\textbf{The Model.---}We focus on a generalization of the SYK model. It consists of $N$ sites with corresponding complex fermions $c_i$, $i \in \lbrace 0, \dots, N \rbrace$ and is described by the following Hamiltonian:
\begin{equation}
\begin{aligned}
\hat{H} &=  \frac{1}{\sqrt{(2N)^3}} \sum_{ijkl}J_{ij;kl} c_i^\dagger c_j^\dagger c_k c_l + \frac{1}{\sqrt{N}}\sum_{ij} f(t) K_{ij} c_i^\dagger c_j,
\end{aligned}
\label{Eq:syk2_syk4}
\end{equation}
where $K_{ij}$ and $J_{ij;kl}$ are chosen to be random, uncorrelated all-to-all couplings which are drawn from Gaussian distributions with means $\overline{K_{ij}},\overline{J_{ij;kl}} = 0$ and variances of $\overline{K_{ij}K_{ij}^*} = |K|^2$ and $\overline{J_{ij;kl}J_{ij;kl}^*}=|J|^2$, respectively. The line denotes a disorder average over realizations of $K_{ij}$ and $J_{ij;kl}$. We drive the system out of equilibrium by modulating the hopping term with a periodic drive $f(t)$ of period $\bar{T} = 2\pi/\Omega$, see 
\fig{fig:schematic}. In equilibrium and at zero hopping, $K_{ij} = 0$, the single-particle correlation function  exhibits a scaling solution $G(t)\sim 1/\sqrt{|{t}|}$ and the fermions acquire an anomalous dimension of $1/4$. In this regime quasiparticles do not exist and the density of states (DOS) diverges as $\sim 1/\sqrt{\omega}$, which is a strong NFL feature. However, in the thermodynamic limit any finite hopping $K_{ij} \neq 0$ constitutes a relevant perturbation, such that the low-energy physics is described by long-lived fermions. Their lifetime is of the usual Fermi liquid form and diverges for low temperatures as $\tau_{\mathrm{LT}} \sim 1/T^2$. Although the system is a FL at the lowest energies, it crosses over to the NFL scaling solutions above the crossover scale $E_c \sim K^2 /J$~\cite{balents}. 

\textbf{Non-equilibrium formalism.---}Using an exact field-theoretic treatment, we solve for the heating dynamics of the SYK model \eqw{Eq:syk2_syk4} as it is periodically modulated in time with a sinusoidal drive of frequency $\Omega$ and amplitude $a$. In the thermodynamic limit $N\rightarrow\infty$ only a small subset of Feynman diagrams contributes, which can be resummed. The Kadanoff Baym equations (KBEs) that govern the evolution of the disorder-averaged Green's function are then given by the following diagrams:

\begin{fmffile}{sde_feynman}
	\begin{align*}
	\begin{gathered}
	\begin{fmfgraph*}(30,20)
	\fmfleft{i}
	\fmfright{o}
	\fmf{fermion}{i,o}
	\end{fmfgraph*}
	\end{gathered}=
	\begin{gathered}
	\begin{fmfgraph*}(30,20)
	\fmfleft{i}
	\fmfright{o}
	\fmf{fermion,fore=(0.7,,0.7,,0.7)}{i,o}
	\end{fmfgraph*}
	\end{gathered}+
	\begin{gathered}
	\begin{fmfgraph}(70,10)
	\fmfleft{i}
	\fmfright{o}
	\fmf{phantom}{i,v1,v2,v3,v4,o}
	\fmf{dots,left=1.5,tension=0.01}{v2,v4}
	\fmf{fermion,fore=(0.7,,0.7,,0.7),tension=3.5}{i,v2}
	\fmf{fermion,tension=3.5}{v4,o}
	\fmf{fermion,left,tension=1}{v2,v4,v2}
	\fmf{fermion}{v2,v4}
	\fmfdot{v2,v4}
	\end{fmfgraph}	
	\end{gathered}+
	\begin{gathered}
	\begin{fmfgraph}(70,20)
	\fmfleft{i}
	\fmfright{o}
	\fmf{phantom}{i,v1,v4,o}
	\fmf{dots,left=1.,tension=0.2}{v1,v4}
	\fmf{fermion,fore=(0.7,,0.7,,0.7),tension=2.}{i,v1}
	\fmf{fermion,tension=2.}{v4,o}
	\fmf{fermion}{v1,v4}
	\fmfdot{v1,v4}
	\end{fmfgraph}	
	\end{gathered},
	\end{align*}
\end{fmffile}where convolutions in time are performed over the Kadanoff-Baym round-trip contour $\mathcal{C}$ going to times infinity and back. Gray lines denote the free Green's function, $G_0$; black lines the disorder averaged full Green's function, $G(t,t') = -i\overline{\langle\mathrm{T}_\mathcal{C}\lbrace c_i(t) c_j^{\dagger}(t')\rbrace  \rangle}$ (where $\mathrm{T}_\mathcal{C}$ denotes contour time ordering); and dotted lines represent disorder averages of $K_{ij}$ and $J_{ij;kl}$, respectively. The precise form of our KBEs is read off from the diagrams to be
\begin{equation}
\begin{aligned}
G(t,t') &= G_0(t,t') + \int_\mathcal{C} d2 d3 \; G_0(t,2)\Sigma(2,3)G(3,t') \\
\Sigma(1,2) &= J^2 \; G(1,2)^2G(2,1) + f(1)f(2)\, K^2\; G(1,2).
\end{aligned}
\label{eq:SchwingerDyson}
\end{equation} 
As initial conditions we fix the Green's function in the lower quadrant of the two-time plane $(t,t')$ with low-temperature initial states, which are determined by numerically iterating the equilibrium Schwinger-Dyson equations~\cite{Maldacena2016Remarks,Babadi2013PhD,Eberlein2017Quench}.

We then switch on the drive at time $t=0$ and follow the evolution of the system. The resulting equations are integrated numerically employing a predictor-corrector scheme; see Supplemental Material for details~\cite{supp}\nocite{rammer_smith}\nocite{Babadi2015Spiral}.
Following this formalism we obtain self-consistent and conserving solutions for the Green's functions arbitrarily far from equilibrium.

\textbf{Heating dynamics of a heavy Fermi liquid.---}
\begin{figure}[t]
\centering
     \includegraphics[width=1\columnwidth]{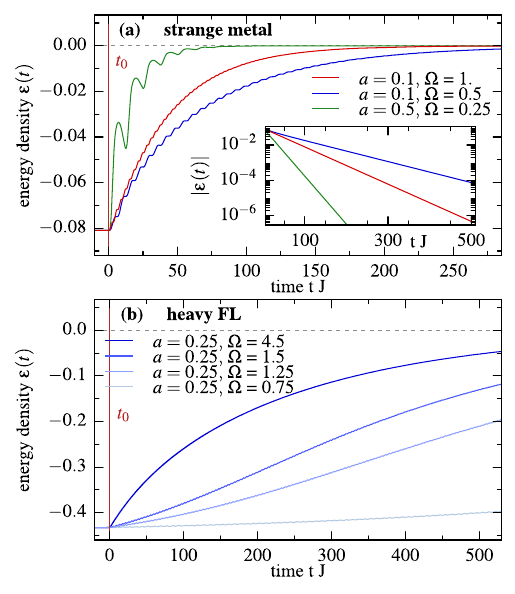}
\caption{\textbf{Energy absorption.} \textbf{(a)} Driving the strange metal leads to a rapid exponential heating as confirmed for a variety of drive parameters. \textbf{(b)}  The driven heavy Fermi liquid displays a crossover from slow to exponential heating when $\Omega \lesssim K^2/J$. By contrast, for $\Omega \gtrsim K^2/J$ the dynamics becomes exponential rather quickly.}\label{fig:2}
\end{figure}
We periodically drive both the NFL (also referred to as \emph{strange metal}) which arises at low temperatures when the hopping is negligible, and the \textit{heavy FL}, which is the relevant low-energy description for finite hopping $K_{ij} \neq 0$. Both cases are expected to be discernible from their short time dynamics, which will be dominated by the low-energy excitations that are vastly different in the two cases. We initially prepare the system in a low-temperature state $T \ll \mathrm{min}(J, K)$, and then compute the absorbed energy as a function of time from a diagrammatic expansion to leading order in $N$,
 \begin{fmffile}{energy_feynman}
	\begin{equation}
	\begin{aligned}
	\epsilon(t)  &= \overline{tr\lbrace \rho_\text{th} H(t)\rbrace} =  \frac{1}{2N}	    \begin{gathered}
	\begin{fmfgraph*}(60,60)
	\fmfleft{i}
	\fmfright{o}
	\fmf{phantom,tension=10}{i,v1}
	\fmf{phantom,tension=10}{v2,o}
	\fmf{dots,left=1.5,tension=0.01}{v1,v2}
	\fmf{fermion,left,tension=0.4}{v1,v2,v1}
	\fmf{fermion,left=0.5}{v1,v2}
	\fmf{fermion,left=0.5}{v2,v1}
	\fmfdot{v1,v2}
	\end{fmfgraph*}
	\end{gathered} 
	+ \frac{1}{N}
	\begin{gathered}
	\begin{fmfgraph*}(60,60)
	\fmfleft{i}
	\fmfright{o}
	\fmf{phantom,tension=10}{i,v1}
	\fmf{phantom,tension=10}{v2,o}
	\fmf{dots,left=1.,tension=0.02}{v1,v2}
	\fmf{phantom,left,tension=0.4}{v1,v2,v1}
	\fmf{fermion,left=0.5}{v1,v2}
	\fmf{fermion,left=0.5}{v2,v1}
	\fmfdot{v1,v2}
	\end{fmfgraph*}
	\end{gathered} 
	\\
	&= -i\int_\mathcal{C}dt'\; \frac{J^2}{2} G^\mathcal{C}(t,t')^2 G^\mathcal{C}(t',t)^2+ K^2 G^\mathcal{C}(t,t') G^\mathcal{C}(t',t),
	\label{dia:energydirac}
	\end{aligned}
	\end{equation}
\end{fmffile}where the first diagram accounts for interactions and the second for the kinetic energy.

We first study the heating of the \textit{strange metal} phase, by setting $f(t) = a \sin(\Omega t)$ in Eq.~\ref{Eq:syk2_syk4}, which corresponds to a drive with zero mean. The system is prepared at low temperatures $\beta J = 50$, which realizes an approximately conformal state. Because of the strong scattering and the absence of quasiparticles the absorbed energy increases exponentially comparatively quickly after switching on the drive and remains so as it crosses over to the free ultraviolet theory, see \figc{fig:2}{a}. This exponential heating is of the form
\begin{align*}
|\epsilon_{\mathrm{SYK}}(t)| \sim e^{- \Gamma(a,\Omega) t}
\end{align*}
and arises with a frequency and amplitude dependent rate $\Gamma(a, \Omega)$. The parameter dependence of $\Gamma(a,\Omega)$ is studied in the Supplemental Material for close to resonant driving. We find a universal scaling with the driving amplitude $\Gamma(a,\Omega)\sim a^2 f(\Omega)$.

The heating of the \textit{heavy Fermi liquid} can be richer. We again prepare the system in a low-temperature state with $\beta J = \beta K = 50$ which lies deep within the FL regime. We drive the system by modulating the hopping term $K_{ij}$ periodically around a mean value,  $f(t) = 1 + a \sin(\Omega t)$. In this case the drive commutes with the hopping, but not with the interaction. Initially, quasiparticles which are excited are long-lived, since $\tau_{\mathrm{FL}} \sim 1/T^2$. For low drive frequency $\Omega$, the system therefore exhibits two heating regimes: (I) subexponential heating at early times and (II) late time exponential heating. The crossover scale for the transition from slow (I) to fast (II) heating is tuned by $\Omega$, Fig.~\ref{fig:2}(b) and is set by $E_c \sim K^2/J$. 
While the absorption at short times depends on the nature of the drive, we find an exponential heating at late times as a universal property.

\textbf{Thermalization.---}To understand the heating of the system beyond the dynamics of the energy density, we also study observables that more directly probe the many-body state. In particular, we will evaluate generalized fluctuation dissipation relations (FDR) to extract effective temperatures. To this end, we transform the Green's function to center of mass coordinates, $t = (t_1+t_2)/2$, and relative coordinates, $\Delta t = t_1-t_2$, and Fourier transform the latter. Subsequently we compare the statistical or Keldysh component $G^K = - i \langle [c(t), c^\dagger(t')]\rangle$ and the spectral components $A = \langle \lbrace c(t), c^\dagger(t')\rbrace\rangle$ of the Green's function as follows
\begin{align}
G^K(t, \omega) = \tanh\left(\frac{\beta_{\mathrm{eff.}}(t)\omega}{2}\right) i A(t,\omega).
\end{align}
Provided that the FDRs are fulfilled at low energies, we can extract the time-dependent effective temperature $T_{\mathrm{eff.}} = 1/\beta_{\mathrm{eff.}}$ to characterize the state. 

We find the FDRs for the \textit{strange metal} to be rapidly fulfilled for comparably weak driving, yielding an effective time dependent temperature $T_{\text{eff.}}(t)$ shortly after the drive is switched on. We can compare this temperature, with a thermal state evaluated at the instantaneous energy density via the self-consistent relation \begin{align}\label{eq:inst}
\epsilon(t) = \langle \hat H\rangle_{T_\text{inst.}(t)}.
\end{align} 
Comparing, the two temperatures $T_{\text{eff.}}(t)$ and $T_\text{inst.}(t)$ tells us how thermal the system is at a certain time $t$. 

\begin{figure}[t]
	\includegraphics[width=1.\columnwidth]{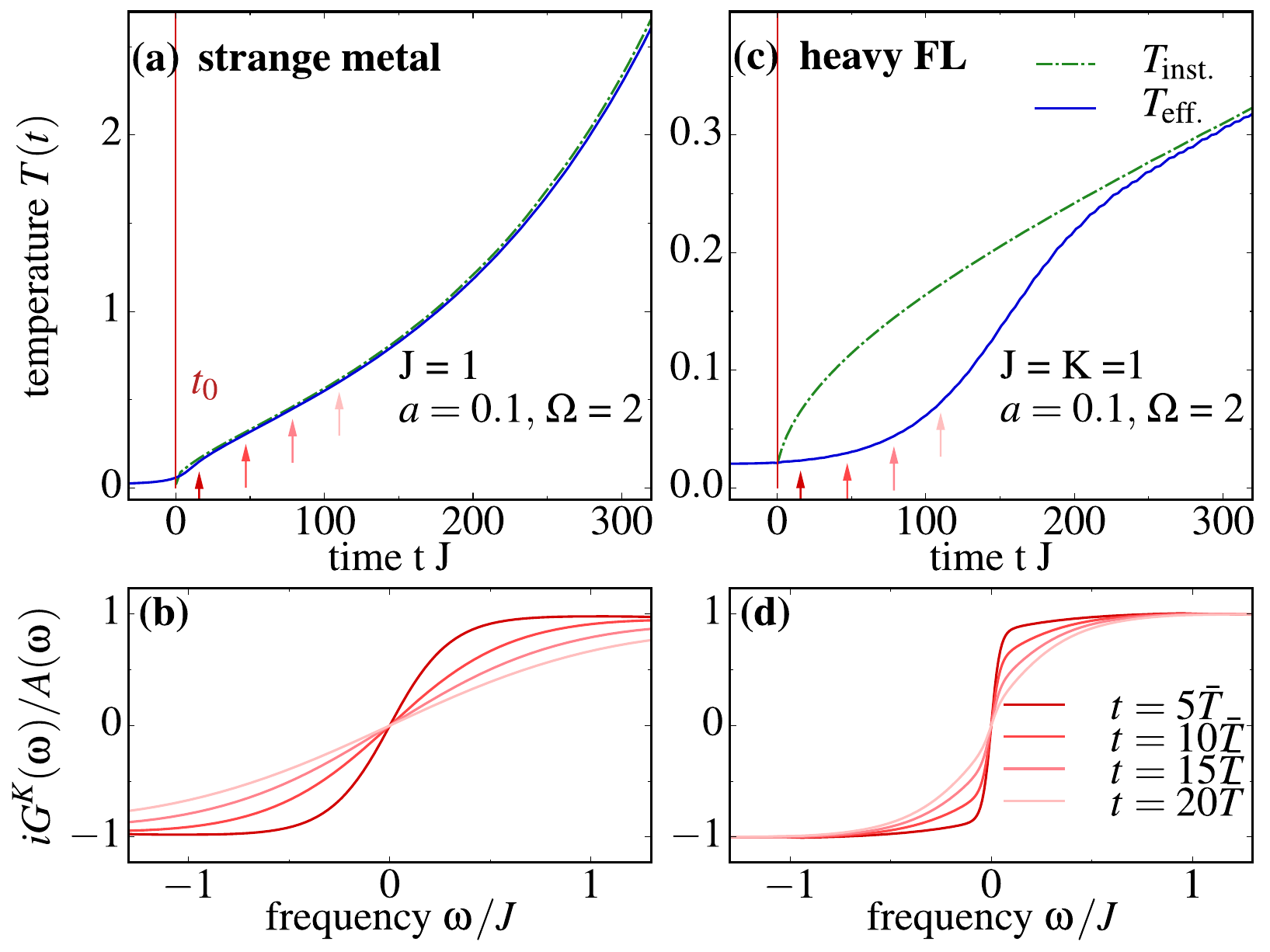}
	\caption{\textbf{Fluctuation-dissipation relations under periodic driving.} \textbf{(a)} For the driven strange metal, the effective temperature $T_\text{eff.}(t)$ obtained from FDRs quickly agrees with the instantaneous temperature $T_\text{inst.}(t)$ obtained from assuming thermal equilibrium at the instantaneous energy density, see \eq{eq:inst}, highlighting the fast thermalization. \textbf{(b)} FDRs at four different stroboscopic times indicated by the arrows in (a).  \textbf{(c)} The Fermi liquid fails to thermalize for long times, indicated by the deviations between the two temperatures. \textbf{(d)} The corresponding FDRs are steplike and are only fulfilled at low energies.}
	\label{fig:fdr}
\end{figure}

The strange metal shows almost instantaneous thermalization, since the two temperatures agree already at short times, see of Fig.~\ref{fig:fdr}(a) and the FDRs are fulfilled over a reasonably large energy regime (b). Thus the system quickly falls back to its equilibrium albeit at a temperature that increases with time. At late times the temperature grows exponentially as $T_{\mathrm{eff.}}^\mathrm{SYK}(t) \sim e^{\Gamma(a, \Omega)t}$. This is connected to the exponential energy absorption by the diverging specific heat $C_V^{\mathrm{SYK}} \sim 1/T^2 + \mathcal{O}(1/T^4)$.  

By contrast, the presence of quasiparticles in the \textit{heavy Fermi liquid} obstructs thermalization on short timescales and FDRs are fulfilled only for small energies, see Fig.~\ref{fig:fdr}(d), where the FDRs exhibit kinks at low frequencies which are reminiscent of the low-temperature initial state. Extracting an effective temperature scale from the low-energy behavior shows that for long time spans the two different notions of temperature strongly disagree, \figc{fig:fdr}{c}. Only at late times the system thermalizes and well-defined FDRs emerge, indicating that the system is in thermal equilibrium. 

For weak driving we can estimate the emergent temperature scale $T^*$ at which the system thermalizes by comparing the dissipative response $\Gamma_E \sim a^2 \Omega \chi''(\Omega)$, where $\chi''(\Omega)$ is the imaginary part of the kinetic energy correlator, with the decay rate of excitations in the heavy FL $1/\tau_\text{LT} \sim T^2$, which gives $T^* \sim a \sqrt{\Omega \chi''(\Omega)}$. The effective thermalization timescale is then set by $t^* \sim 1/T^*$ and is thus inversely proportional to the driving strength. The linear scaling of $T^*$ is confirmed by our exact numerical simulations in the Supplemental Material. By contrast in the strange metal the inverse life-time scales as $1/\tau^\text{NFL}_\text{LT} \sim T$, leading to a $T^* \sim a^2\Omega \chi''(\Omega)$ that is strongly suppressed at small drive amplitudes. The temperature of our initial state is therefore already of the same order of magnitude as $T^*$ supporting the rapid thermalization of the NFL. Since our model of Eq.~\ref{Eq:syk2_syk4} is rather generic, we expect that slow thermalization can also be present in more complex physical systems with weakly interacting quasiparticles, providing an alternative route of controlling heating dynamics in Floquet systems.

\textbf{High frequency dynamics.---}
\begin{figure}
  \centering
    \includegraphics[width=1\columnwidth]{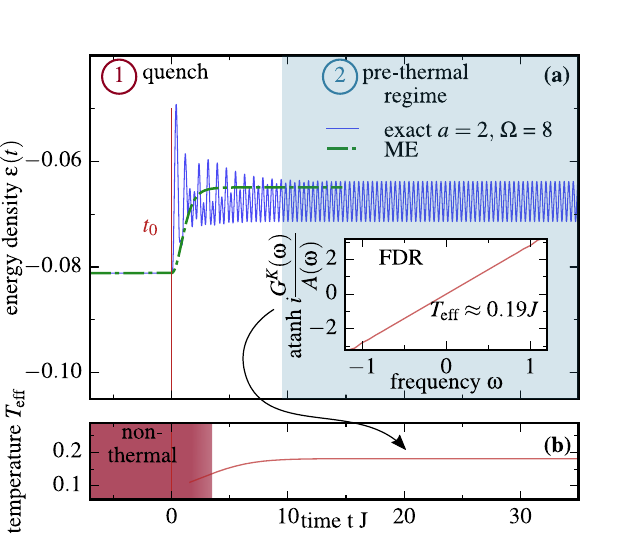}
  \caption{\textbf{Fast driving.} \textbf{(a)} The exact energy density of the system (blue) shows good agreement with the evolution of an effective Floquet Hamiltonian (green dash-dotted line). Turning on the rapid drive acts as a quantum quench, after which the system effectively stops absorbing energy and enters a prethermal plateau.  \textbf{(b)} In this prethermal regime the system is thermal, see FDRs in the inset, with a temperature that is nearly constant but slowly increasing. Only after long times the system can heat up to infinite temperature.}
  \label{fig:magnus}
\end{figure}So far, we have considered drive frequencies that are slow compared to the bandwidth of the couplings. Heating can be slowed down by increasing the driving frequency $\Omega$. The intermediate-time, stroboscopic, evolution of the system is governed by the Floquet Hamiltonian, $H_\mathrm{F} := i\ln[U(\bar{T})]/\bar{T}$, where $U(T)$ is the unitary time evolution operator over one driving period. $H_\mathrm{F}$ can be determined perturbatively in the inverse frequency, $1/\Omega$, of the drive using the Magnus expansion (ME)
\begin{equation}
\begin{aligned}
H_\mathrm{F} &= H_\mathrm{F}^0 + H_\mathrm{F}^1 + \dots,
\end{aligned}
\label{eq:magnus_exp}
\end{equation} 
with $H_\mathrm{F}^0 =\frac{1}{\bar{T}} \int_{0}^{\bar{T}} H(t)$ and $H_\mathrm{F}^1 = \frac{1}{2! \bar{T} i} \int_{0}^{\bar{T}}dt_1 \int_{t_0}^{t_1} dt_2 [H(t_1), H(t_2)]$. This effective high-frequency Hamiltonian is conserved up to exponentially long times $t^* \sim e^{\mathcal{O}(\Omega/J)}$ in $\Omega$. For times $ t \ll t^*$ the system is therefore expected to be approximately static and well described by the ME Hamiltonian.

We find that the fast drive simply renormalizes the coupling $J_{ij;kl}$ of the SYK model. To first order, the ME yields the new effective Hamiltonian:
\begin{equation}
\begin{aligned}
\hat{H}_\mathrm{F} =& \frac{1}{\sqrt{2N}^3}\sum_{ijkl} J_{ij;kl} c_i^\dagger c_j^\dagger c_k c_l  +\\
&+ \frac{-i}{\Omega} \frac{1}{\sqrt{2^3 N^4}} \sum_{ijkl} \tilde{J}_{ijkl} c_i^\dagger c_j^\dagger c_k c_l + \; \mathcal{O}(1/\Omega^2),
\label{eq:floquet_hamilt_syk}
\end{aligned}
\end{equation}
where $\tilde{J}_{ijkl}$ denote the interactions induced by the drive, which are evaluated by explicitly using commutation relations. We find the following couplings
$\tilde{J}_{ijkl} = 2 \sum_\mu K_{i \mu} J_{\mu j k l} - 2 \sum_\mu J_{ijk\mu} K_{\mu l}$. To this order these couplings can be treated as independent from the original $J_{ijkl}$ due to disorder averages over $K_{ij}$. As the $\tilde{J}_{ijkl}$ are sums over products of Gaussian distributed variables, $\tilde{J}_{ijkl}$ are again Gaussian distributed, by the central limit theorem with mean $\overline{\tilde{J}_{ij;kl}} = 0$ and variance  $\overline{\tilde{J}_{ij;kl}\tilde{J}_{ij;kl}^*}=4 |J|^2 |K|^2$.

Starting from a low-temperature state $\beta J = 50$ and switching on a rapidly oscillating drive, we find that the dynamics is divided into two regimes (Fig.~\ref{fig:magnus}): (I) At short times, the rapid drive acts as a perturbation that rearranges the many-body state and leads to some finite energy absorption. The stroboscopic dynamics is approximately described by a quantum quench to the Floquet Hamiltonian, $\hat{H}_\mathrm{F}$. (II) After that the energy absorption is strongly suppressed and the system enters a prethermal regime. By solving for the exact dynamics we find that the system respects effective FDRs, see \figc{fig:magnus}{a} inset. In the prethermal phase the system realizes a NFL with a rapid relaxation rate given by $1/\tau_{\text{NFL}} \sim T$. These fast scattering processes have to be compared to the weak absorption, which we find to be superexponentially suppressed in the driving frequency (see Supplemental Material). This separation of scales guarantees the emergence of long-lived prethermal dynamics and approximate FDRs. We have confirmed that the effective temperature increases steadily, albeit at a strongly suppressed rate. Although the system slowly heats to infinite temperature, it remains trapped in a quasistatic strongly correlated state for long times and evolves according to the effective Floquet Hamiltonian.
 
\textbf{Graphene flakes.---}Graphene flakes with sufficiently disordered boundaries, in strong background magnetic fields have recently been proposed to realize exotic SYK physics~\cite{ChenGrapheneFlake}. Here, we generalize these ideas to include driving by classical laser light that impinges onto the graphene flake; see Supplemental Material~\cite{supp}. This leads to a  periodically driven SYK Hamiltonian as in \eq{Eq:syk2_syk4}, where $N$ is proportional to the number of zero modes in the lowest Landau level which scales with the flux through the sample. As the strange metal behavior appears for $N \gg 1$, high magnetic fields are required. The flux can, however, also be increased by considering larger flakes that contain irregularly shaped holes in their interior~\cite{atac}, allowing one to obtain the same number of modes $N$ in the lowest Landau level albeit at smaller applied magnetic fields. The light-induced driving term then dissipates energy in the system and leads to heating, which can, for example, be measured by transmission spectroscopy. To qualitatively predict the behavior of the dynamics, one needs to study the disorder parameters for realized flake geometries. 

\textbf{Conclusions and Outlook.---}We studied the heating dynamics in periodically driven heavy Fermi liquids and strange metals. We focus on generalized SYK models, which remain exactly solvable using non-equilibrium field theory. These systems absorb energy exponentially quickly at late times, as they approach their infinite temperature states. While the strange metal thermalizes almost instantaneously, the heavy Fermi liquid requires a comparatively long time to equilibrate.  We argue that this behavior is a result of relatively weakly interacting quasiparticle excitations whose lifetime scales as $1/T^2$ at low temperatures.  This effect provides an alternative route to suppress heating in more complex periodically driven systems in the future. We also report a prethermal regime for high-frequency drives and propose experimental opportunities to probe non-equilibrium SYK physics irradiating irregularly shaped graphene flakes.

Our approach can be readily generalized to higher-dimensional SYK systems with potentially topologically nontrivial band structures. Such models would offer new routes to explore the interplay of transport topology, periodic driving and strong correlations.

\begin{acknowledgments}
\textit{\textbf{Acknowledgments.---}}We thank I. Esin, J. Feldmeier, A. Imamoglu, N. Lindner, A. Rosch, S. Weidinger and H. Weyer for interesting discussions. The authors acknowledge support from the Technical University of Munich - Institute for Advanced Study, funded by the German Excellence Initiative and the European Union FP7 under grant agreement 291763, the Deutsche Forschungsgemeinschaft (DFG, German Research Foundation) under Germany's Excellence Strategy--EXC-2111--390814868, the European Research Council (ERC) under the European Union's Horizon 2020 research and innovation programme (grant agreement No. 851161), from DFG grants No. KN1254/1-1, KN1254/1-2 and DFG Projekt-ID 107745057 -- TRR 80.
\end{acknowledgments}

\appendix
\newpage
\setcounter{figure}{0}
\setcounter{equation}{0}

\renewcommand{\thepage}{S\arabic{page}} 
\renewcommand{\thesection}{S\arabic{section}} 
\renewcommand{\thetable}{S\arabic{table}}  
\renewcommand{\thefigure}{S\arabic{figure}} 
\renewcommand{\theequation}{S\arabic{equation}} 

\onecolumngrid

\begin{center}
\textbf{\Large{\large{Supplemental Material: \\The periodically driven Sachdev-Ye-Kitaev model}}}
\end{center}
\subsection{Schwinger Dyson equations on the Kelydsh contour}\label{sec:SDE}
We  obtain the non-equilibrium correlation functions of our system by solving the exact Schwinger Dyson equations on a Keldysh contour. Here we sketch the iterative numerical solution to the equations. The Schwinger Dyson equations in real time are a set of integro-differential equations, which are derived by acting with the free inverse Green's function $G_0^\mathcal{C}(t,t')^{-1}= i\partial_t i \delta^\mathcal{C}(t-t')$ in Eq.~(2). Out of equilibrium one defines the greater and lesser component of the Green's function to be
\begin{equation}
\begin{aligned}
G^>(t,t') = -i \langle c(t) c^\dagger(t')\rangle, \qquad
G^<(t,t') = i \langle c(t')^\dagger c(t) \rangle.
\end{aligned}
\label{eq:greater_lesser}
\end{equation}
Contour functions and their convolutions can be decomposed into these components by applying the Langreth rules~\cite{rammer_smith}. This decomposition leads to the following evolution equations for the correlation functions
\begin{equation}
i\frac{d}{dt} G^\lessgtr(t,t') = \int_{t_0}^{max(t,t')} d\tau \Sigma^R(t,\tau) G^\lessgtr(\tau,t') + \Sigma^\lessgtr(t,\tau) G^A(\tau,t').
\label{eq:kadbaym}
\end{equation}
We integrate Eq.~\ref{eq:kadbaym} using the trapezoid rule and obtain the prescription
\begin{equation}
	G^\lessgtr(T+\Delta t,t') = G^\lessgtr(T,t') - i\frac{\Delta t}{2} \left[ \int_{-\infty}^{max(T,t')} dt''  \; \Sigma^R G^\lessgtr + \Sigma^\lessgtr G^A +  \int_{-\infty}^{max(T+\Delta t,t')} dt''  \; \Sigma^R G^\lessgtr + \Sigma^\lessgtr G^A \right].
	\label{eq:SchwingerDysonnumerics}
\end{equation}
As both sides explicitly depend on $G^\lessgtr(T+\Delta t)$, we employ a predictor corrector scheme following Ref.~\onlinecite{Babadi2015Spiral}. Step by step the procedure is implemented as follows:
\begin{enumerate}[(I)]
	\item Start with some initial Green's functions $G^{\lessgtr}(t_1,t_2)$ defined on a subset of the $t_1-t_2$ plane $\lbrace (t_1,t_2) | -\infty < t_{1,2} \leq T \rbrace$ which characterizes the (thermal) initial state
	\item Predict the entries of the Green's functions $G^\lessgtr_p(T+\Delta t) $ along the line $t'\in (-\infty, T]$ for some arbitrary quench protocol by setting $G^\lessgtr(T+\Delta t) \approx G^\lessgtr(T,t')$ on the right hand side of Eq.~\ref{eq:SchwingerDysonnumerics}
	\item  Calculate the predicted self energies $\Sigma^\lessgtr_p = \Sigma^\lessgtr[G_p(T+\Delta t, t')]$
	\item Determine the corrected entries of the Green's functions $G^\lessgtr_c(T+\Delta t)$ by setting $G^\lessgtr(T+\Delta t) = G^\lessgtr_p(T+\Delta t)$ on the right hand side of the evolution equation.
\end{enumerate}
This procedure is iterated until $|G^\lessgtr_c - G^\lessgtr_p| < \delta$ converged below some tolerance. $\delta$, $G^\lessgtr(T+\delta t)$ is then set equal to $G^\lessgtr_c$ and we continue by predicting the next time slice. For our purposes we chose a typical grid-size of the $t_1-t_2$ plane of $8000 \times 8000$ and set $\delta = 10^{-8}$. The initial conditions are found by iterating the SDE in frequency space~\cite{Maldacena2016Remarks,Eberlein2017Quench}.

\subsection{Graphene flakes}
Here we study the effect of laser light impinging on an irregularly shaped graphene flake by following the construction of Ref.~\onlinecite{ChenGrapheneFlake}. As discussed in the main text, such a setup may provide an experimental realization of the model in Eq.~(1).
\texttt{\begin{figure}[t]
	\includegraphics[width=.35\textwidth]{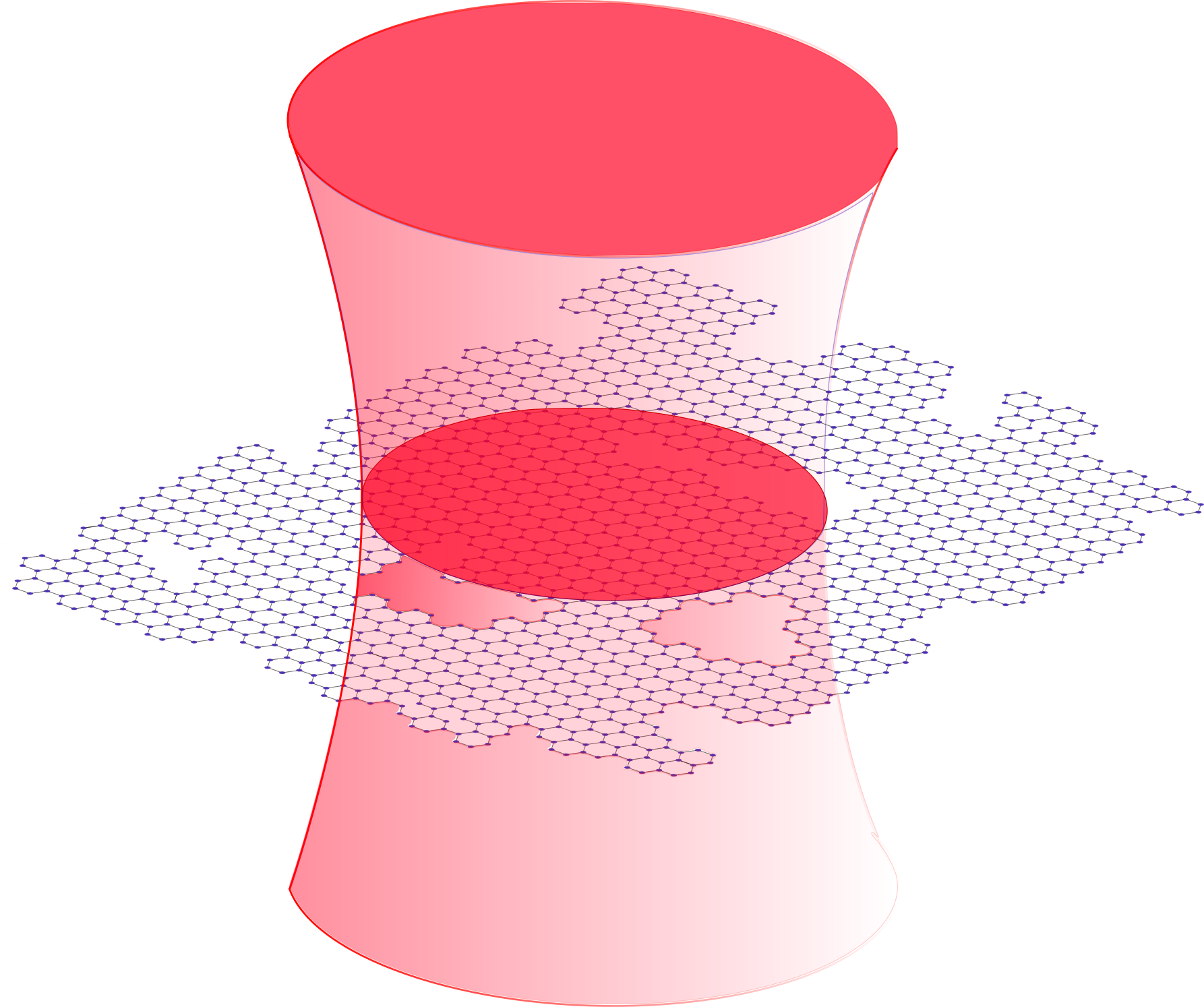}
	\caption{\textbf{Floquet dynamics in graphene flakes.} Schematic of laser light impinging on an irregularly shaped graphene flake. The laser light induces a periodically modulated hopping term in the SYK-description (Eq.~\ref{eq:HamiltonianSYK}). To realize our periodic driving protocol with frequencies $\Omega \sim J$ the laser frequency is expected to lie in the THz regime. The energy density may be determined by measuring the transmission of the beam through the sample.}
	\label{fig:timetraces}
\end{figure}}
The classical light impinging on the flake (see Fig.~\ref{fig:timetraces}) is modelled by a Peierls substitution in the underlying tight-binding model. Introducing a screened Coulomb interaction between the electrons, the Hamiltonian of the graphene flake takes the form
\begin{equation}
\hat{H}_{\text{lab}} = -J \sum_{\langle ij \rangle} e^{i\vec{A}(t)(\vec{r}_i -\vec{r}_j)} c_i^\dagger c_j + \frac{U}{2} \sum \hat{n}_i V(i-j) \hat{n}_j,
\label{eq:hamiltonian_labframe}
\end{equation}
where $\vec{A} = (\cos(\Omega t)- 1/2 B y, \cos(\Omega t) + 1/2 B x, 0)^T$ is the vector potential for the light field with frequency $\Omega$ and $B$ is the background magnetic field. $V(i-j)$ is the potential for density-density interactions. It is advantageous to remove the light-induced hopping amplitude in favour of a time-dependent modulation of the local density. As in Ref.~\cite{BukovGopalakrishnan2015} this is achieved by transforming to a rotating frame by means of a unitary
\begin{align*}
V(t) = e^{-i \vec{A}_{light}(t) \sum \vec{r}_i \hat{n}_i}.
\end{align*}
In the rotating frame all expectation values remain unchanged and the time-evolution is generated by the transformed Hamiltonian
\begin{equation}
\begin{aligned}
H_\text{rot} = \partial_t\vec{A}(t) \sum\vec{r}_i \hat{n_i} -J \sum_{\langle ij \rangle} e^{i\vec{A}'(\vec{r}_i -\vec{r}_j)} c_i^\dagger c_j + \frac{1}{2} \sum \hat{n}_i V(|i-j|) \hat{n}_j,
\end{aligned}
\end{equation}
now with a constant vector potential $\vec{A}' = (-1/2 B y, 1/2 Bx, 0)^T $. By projecting out all but the lowest Landau level (LLL) and by switching to the eigenbasis of the non-interacting problem, the flake is mapped onto a charged SYK Hamiltonian
\begin{equation}
\begin{aligned}
H^\text{LLL}(t) = \frac{1}{2!} \sum_{ijkl=1}^{N} J_{ij;kl}\; c_i^\dagger c_j^\dagger c_k c_l +\sum_{\alpha \beta=1}^{N} F_{\alpha\beta} c_\alpha^\dagger c_\beta  +f(t) \sum_{\alpha \beta=1}^{N} A_{\alpha\beta} c_\alpha^\dagger c_\beta,
\end{aligned}
\label{eq:HamiltonianSYK}
\end{equation} 
where $J_{ij;kl}$, $A_{\alpha,\beta}$ and $F_{\alpha\beta}$ are random all-to-all couplings. They are approximately gaussian distributed, with zero mean and variances $\overline{|J_{ij;kl}|^2} = \frac{J^2}{N^3}$, $\overline{|A_{\alpha\beta}|^2} = \frac{A^2}{N}$ and $\overline{|F_{\alpha\beta}|^2} = \frac{F^2}{N}$~\cite{ChenGrapheneFlake}. $N$ is proportional to the number of zero modes in the LLL and scales with the flux through the sample. The strange metal behavior appears as $N \gg 1$ such that large magnetic fields are necessary. As discussed in the main text, the flux may also be increased by considering larger flakes which contain disordered interiors~\cite{atac}. The light-induced driving term dissipates energy in the system and leads to heating. This may realize an experimental probe of non-equilibrium dynamics in the SYK model. The energy density is accessible by transmission spectroscopy. The heating dynamics could indicate whether the system realizes a clean SYK model, or whether quadratic perturbations dominate the dynamics. While the phases are always distinguishable by studying the thermal properties of the system, FL and NFL cannot be discerned by studying the energy absorption alone if drive and hopping terms would turn out to be completely uncorrelated. In this case both phases would heat exponentially. As the correlations between $F_{ij}$ and $A_{ij}$ determine the functional form of the heating, they need to be analyzed on a concrete flake geometry.

\subsection{Thermalization timescale}
Floquet engineering exotic phases of matter often relies on the equilibration of the modulated system, as the dynamics of periodically driven many-body systems can be complex. This thermalization is well understood in the asymptotic limit of fast driving, where the system is effectively static and heating is strongly suppressed. Alternatively, when a system relaxes sufficiently quickly it remains close to an equilibrium state at all times. Such approximations can be justified above a crossover temperature $T^*$, which emerges dynamically and determines the onset of thermalization in Floquet systems~\cite{rosch}. Here we analyze this scale in detail for a heavy Fermi liquid. Our numerics show a linear dependence of $T^*$ on the driving amplitude $a$ as long as $a$ is small, Fig.~\ref{fig:thermscale}. The scale $T^*$ is determined as the lowest temperature of the driven system at which well-defined FDRs are found. As discussed in the main text, this result is explained as a competition of relaxation and energy dissipation. Generically energy is dissipated at a rate $1/\tau_{\text{diss.}}\sim \Omega f(\Omega) a^2$, while scattering of quasiparticles scales as $1/\tau_{\text{scatter}}\sim T^2$. In Fermi liquids one therefore finds $T^*\sim a$. As scattering in NFL's is a much faster process which scales $\sim T$,  one finds $T^*_{\text{NFL}}\sim a^2$, which is consistent with the rapid thermalization of the NFL phase found in the main text. Corrections to the linear behavior arise due to non-linear absorption when $a$ becomes large, and once the relaxation rate depends significantly on the frequency of the excitations induced by the drive.
\texttt{\begin{figure}[h]
	\includegraphics[width=0.8\textwidth]{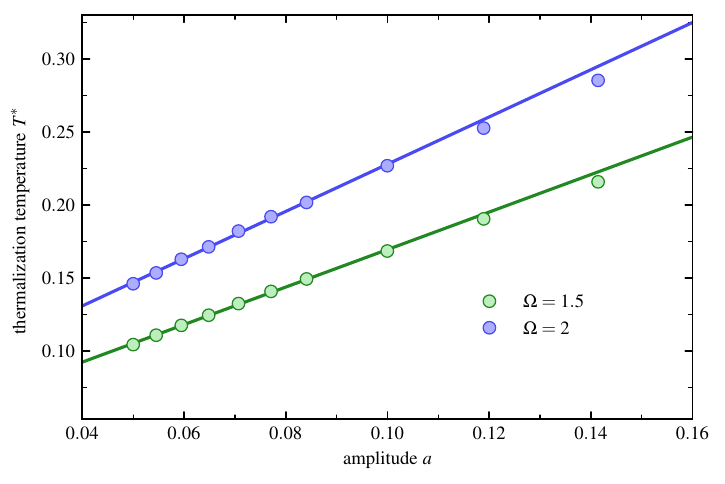}
	\caption{\textbf{Thermalization of a periodically modulated heavy Fermi liquid.} 
Competition of relaxation and energy dissipation sets a thermalization scale $T^*$, which grows approximately linearly in the driving amplitude $a$. We plot $T^*$ as a function of driving amplitude for two driving frequencies $\Omega$. We determine $T^*$ by the onset of well-defined effective fluctuation-dissipation relations (markers). Solid lines represent linear fits to our data up to $a\approx 0.1J$.}\label{fig:thermscale}\end{figure}}
\subsection{Heating rates in linear response theory}
\texttt{\begin{figure}[h]
	\includegraphics[width=0.8\textwidth]{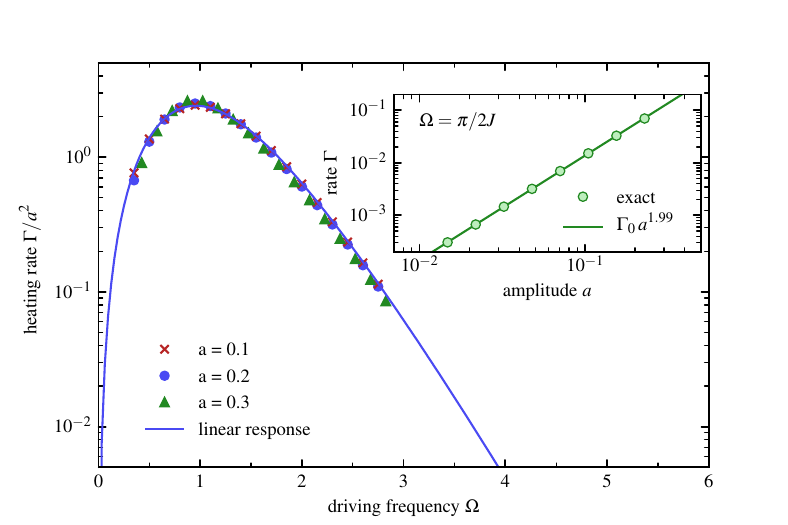}
	\caption{\textbf{Late time heating in the SYK model.} For sufficiently strong absorption, the SYK model heats exponentially with rate $\Gamma(\Omega, a)$. We show the high temperature heating rate as a function of driving frequency (main plot). Markers are obtained by exponential fits to our exact numerical data, while the solid line denotes the linear-response prediction. For high frequencies the rate is super-exponentially suppressed, which signals the emergence of prethermal dynamics. We find that the rate universally scales as $\Gamma(\Omega, a) = F(\Omega) \, a^2$ (inset).}
	\label{fig:linres}
\end{figure}}
The the NFL phase absorbs energy exponentially  $\epsilon(t) \sim \exp(-\Gamma t)$ for a wide regime of drives, where $\Gamma(\Omega, a)$ is a drive-dependent rate. In these cases the heating is almost independent on the initial state, despite being a dynamical property of a strongly interacting system. In the following we study $\Gamma(\Omega, a)$ as a function of drive parameters and demonstrate a universal scaling of $\Gamma\sim a^2$. Fig.~\ref{fig:linres} shows the collapse of rates after rescaling $\Gamma(\Omega,a)$ by $a^{-2}$. The frequency dependence is fixed by a non-universal function $f(\Omega)$. We estimate the rates within a linear response analysis around a high-temperature state and compare to the numerically exact results. To leading order in the driving strength, the absorbed energy per driving cycle is then given by the dissipative part of the response function: $\overline{\mathrm{d} E/\mathrm{d}t} = a^2 \Omega \, \chi''(\Omega)$, which is determined by the commutator $D(t,t') = -i \Theta(t-t') \langle \psi| [H(t), H_{\text{drive}}(t')] |\psi\rangle$. In the SYK model the susceptibility factorizes into equilibrium two-point functions $D(t) \sim G(t)G(-t)$ to leading order in $N$, which we determine by numerically iterating the SDEs. We find that linear response theory accurately describes the heating of the NFL phase.
For fast drives $\Omega \gg J$, the system is quenched and rests in a long-lived prethermal NFL state. This regime emerges due to strongly suppressed heating and will not be captured by an expansion around a high temperature state. Accordingly, our linear response analysis shows a super-exponential suppression of $\Gamma$ in this regime. This result falls well within exponential bounds on the energy absorption $ \dot{E}/N \leq \, C e^{- |\kappa|\Omega}$, where $C$ and $\kappa$ are system dependent constants~\cite{AbaninHeating, HeatingBound}.

\end{document}